# Nanomechanical Characterization of the Interfacial Properties of Bilayers of 1,2-dipalmitoyl-sn-glycero-3-phosphocholine.

*Luca Quaroni\*, Kingshuk Bandopadhyay, Marzena Mach, Paweł Wydro, Szczepan Zapotoczny*

*Department of Physical Chemistry and Electrochemistry, Faculty of Chemistry, Jagiellonian University, ul. Gronostajowa 2, 30-387, Cracow, Poland*

*E-mail: luca.quaroni@uj.edu.pl*

**Abstract**

We investigate the mechanical properties of bilayers of 1,2-dipalmitoyl-sn-glycero-3-phosphocholine in the gel phase by using peak force tapping with quantitative nanomechanical mapping. We study both dry and aqueous bilayers and liposomes supported on the oxidized silicon surface. We report the observation of a marked substrate effect on the measured Young modulus of supported bilayer stacks which decreases as the height of the stack increases. In contrast a clear substrate effect is not observed for the top bilayer of supported aqueous liposomes, which is however affected by the surface curvature of the sample. Adhesion forces present quantitative differences between dry and aqueous samples, with the former being dominated by capillary effects and the latter by non-contact interactions between tip and substrate. The mechanical properties of stacked bilayers reveal a threshold between two different regimes, for the lower portion and the upper portion of the stack, that reveals a change in the plasticity of the system. The threshold is affected by applied setpoint and aging of the sample. We propose that it arises from the presence of thin layers of hydration water between the headgroups of contacting phospholipid bilayers, which appears to have a major effect on the response of the bilayer to an external mechanical stress.

**Introduction**

Phospholipid bilayers are lyotropic liquid crystalline structures subject to investigation in a variety of scientific fields. At one end of this spectrum, when comprised of a small number of molecular components, they form highly organized systems which fall within the domain of interest of material science, physics and physical chemistry. In their most complex embodiment, in conjunction with proteins and other biomolecules, they constitute the membrane scaffold that organizes the structure and dynamics of living cells.[1] In all the systems under investigation, the mechanical properties of the bilayer play a fundamental part in the understanding of structure function relationships. A detailed quantitative study of the mechanics of phospholipid bilayer properties is generally unsatisfactory when performed on macroscopic samples, because it fails to account for the effect of their anisotropic structure at the level of the single molecule, requiring the use of techniques that fall into the domain of nanotechnology.

The introduction of the atomic force microscope (AFM) [2] has rapidly led to its application to the study of mechanical properties of microscopic and nanoscopic structures. The desired quantities are extracted from force-distance curves by using theoretical models that relate tip indentation to the force exerted on the sample. The analysis of a single curve can in principle provide such properties with a spatial resolution of the order of the tip-sample contact. The introduction of Peak Force (PF) Tapping has opened the possibility to perform such analysis in real time, providing nanomechanical maps recorded in synchrony with topography. In a PF measurement the tip engages the sample with a sinusoidal motion until a set value of the applied force is reached and disengages. The whole cycle is repeated for each spatial location while the tip is performing a raster scan of the sample. Each PF tapping cycle effectively corresponds to the engage and disengage sequence of a force curve, which allows extracting mechanical properties of the sample in each single location. Automated analysis of the curve in Figure 1B provides the quantities that constitute the output of a peak force quantitative nanomechanical (PF QNM) measurement.

Figure 1 provides a schematic representation of a PF measurement and its analysis. Figure 1A show the time course of the force applied on the sample as the tip approaches, engages and disengages. In the engagement phase (blue line) the tip approaches the region where attractive van der Waals force are acting and snaps onto the sample. The approach proceeds until the interaction becomes repulsive. The repulsive force increases until reaching the maximum applied force, the peak force, as defined by the setpoint of the experiment. While disengaging (red line) the tip experiences a decrease of repulsive interaction until it reaches a region where force becomes negative as it is retained by adhesive forces, before being released. Figure 1B shows the same cycle in the force versus position curve. The distance on the X axis between the zero-force point during disengagement and the peak force position provides the deformation, or indentation, of the sample. The minimum force of the disengage curve corresponds to the adhesion force. The area between the engage and disengage curves provides the energy dissipated during the cycle because of the plastic deformation of the sample. The disengagement curve, in conjunction with the indentation and the geometrical parameters of the tip, is used to extract the modulus of the sample.

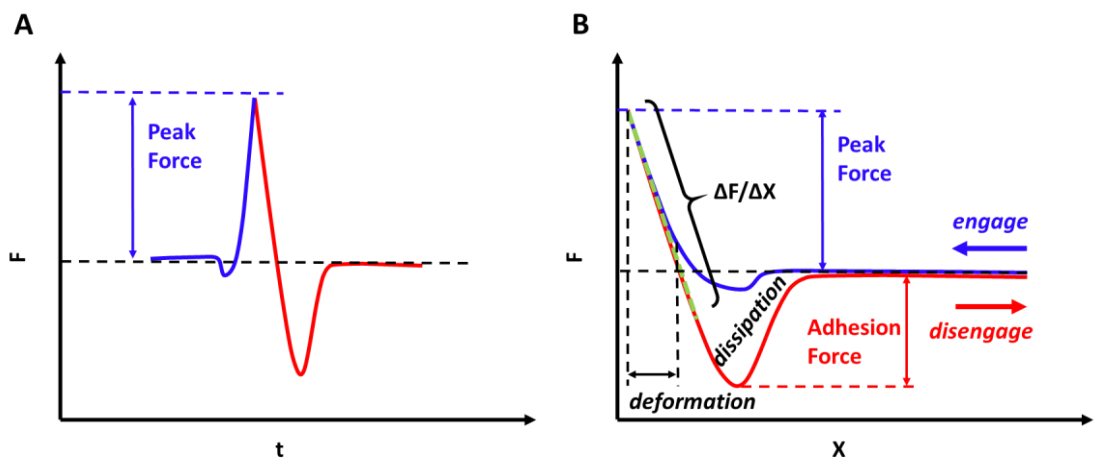

*Figure 1. Measurement of mechanical properties from the Peak Force tapping cycle. **A.** Force vs. time cycle in peak force tapping. **B.** Mechanical quantities extracted from force-distance curves represented in a force vs. tip position plot.*

While generally successful, the accuracy of indentation models relies on the implicit assumption that samples can be described by macroscopic mechanical properties. To satisfy such condition, the material must have infinite thickness, relative to tip size, and be continuous, *i.e.* without the discontinuity implicit in atomic structure. Ultra-thin layers, including phospholipid bilayers and biological membranes, fail to satisfy both these conditions, being only a few nanometers and a few molecules in thickness. It has been shown that for such samples, measured mechanical properties, in particular Young's modulus, are not independent from the properties of the substrate.[3]

Stacked phospholipid bilayers provide an opportunity to investigate the effect of the substrate in a systematic way. The thickness of such systems can be controlled with a discrete progression by increasing the number of bilayers, providing samples that extend from less than 5 nm to more than 100 nm. This range of scales allows us to probe the transition between the mechanical properties of monomolecular layers and those of bulk samples. At the same time, the stack of bilayers retains molecular-scale discontinuities at the level of the phospholipid headgroups and their interfaces. This structure provides for a unique architecture with the simultaneous interplay of atomic and bulk properties.

In the present work we perform a quantitative analysis of the nano-mechanical properties of supported 1,2-dipalmitoyl-sn-glycero-3-phosphocholine (dipalmitoylphosphatidylcholine DPPC) layers, including lamellar multibilayers and vesicles, to assess the relationship between sample geometry and substrate properties in defining the mechanics of the composite system.

**Experimental**

A Dimension Icon atomic force microscope (Bruker, Santa Barbara, CA, USA) working in PeakForce Quantitative Nanomechanical Mapping (PF QNM) mode was used for collecting images in air and in aqueous environment. Bruker Scan-Asyst Water tips were used for all the experiments after calibration. Images of the mechanical properties of the sample were obtained by real time analysis of force curves. Images of Young's modulus of the sample, E, were obtained by using the Derjaguin–Muller–Toporov (DMT) model for indentation. Values of the modulus were also calculated in selected locations of the sample by extracting force curves and manually fitting them using Hertz's and Sneddon's models. The value of the Poisson ratio used for the calculation was 0.3. Data processing and plotting was performed using the freeware package Gwyddion 2.53 (http://gwyddion.net/) and OriginPro 2019 (OriginLab Corp., Northampton, MA, USA).

p-doped (B) Si slides were cleaned by submerging in piranha solution (1 part 30% $H_2O_2$ and 2 parts concentrated sulfuric acid) for thirty minutes. The slides were then thoroughly washed with MQ water and stored under freshly produced MQ water until used for liposome deposition. The piranha cleaned slides display a thin layer of surface silicon oxide and are referred to as $Si/SiO_x$.

1,2-dipalmitoyl -sn-glycero-3-phosphocholine (DPPC) was purchased from Avanti Polar Lipids Inc. Cholesterol (Chol) were supplied by Sigma-Aldrich. All the compounds were the products of high purity (>99%). The solution of DPPC was prepared in chloroform and methanol (9:1 v/v). All the solvents (HPLC grade, ≥99.9%) were obtained from POCH.

Unilamellar liposomes in buffer were prepared by hydration of the dry lipid film, followed by extrusion. The solvent was evaporated under a gentle stream of nitrogen. The dry lipid film was then hydrated with 3 ml of Tris 10 mM/NaCl 150 mM in the presence of 3 mM CaCl2 (pH 7.4 ± 0.1). The resulting multilamellar vesicle dispersion was subjected to five freeze – thaw cycles from liquid nitrogen temperature to about 60°C. Then the dispersion was extruded 13 times through a polycarbonate filter (Whatman) with 100-nm pore size, using a manual extruder purchased from Avanti Polar Lipids. The liposome suspension was further diluted in the same solution to reach a final concentration of lipids 0.1 mg/mL.

Multilamellar liposomes in water were prepared by hydration of the dry lipid film of DPPC with MQ water followed by vortexing and less than 1 min sonication. The resulting multilamellar vesicle dispersion was subjected to five freeze – thaw cycles from liquid nitrogen temperature to about 60°C. The liposome suspension was used without further dilution.

Supported lipids bilayers (SLB) and liposomes in aqueous buffer were prepared via the vesicle fusion method. Briefly, the silicone slides, previously cleaned with a piranha solution, were placed in a weighing vessel and then the suspension of unilamellar liposomes was added to cover the entire slide. The samples were incubated at temperature above phase transition of the lipid bilayer for 60 min and then progressively cooled down to 25 °C over 1 h to ensure full relaxation of the bilayer. After incubation, the samples were washed carefully 5 times with measurement solution (Tris 10 mM/NaCl 150 mM without CaCl2 pH 7.4 ± 0.1) to remove excess vesicles. That was done by creating liquid flow in the solution (pipetting the solution up and down 4 times) and substitute with fresh measurement solution after each wash. The entire washing process was carried out so as not to expose the surface of the slides, and thus not to destroy SLB. The samples were used for AFM measurements in aqueous environment without additional modification and within one day.

Dry bilayer stacks of DPPC were prepared via the vesicle fusion method using multilamellar vesicles in water, without any buffer. After about 1 hour incubation at room temperature, the samples were allowed to dry in air and measured within a few hours.

Results

The dried sample of multilamellar DPPC liposomes was screened to localize larger multibilayer stacks with regular plateaus, which were then characterized using peak force QNM mode. Figure 2 shows the results for an extended multibilayer stack. Analysis of the topography of the sample using the Z Sensor height reading (Figure 2A) shows that the mapped portion displays a terraced structure of stacked bilayers with progressively smaller dimensions.

All terraced levels, except for the first one, have a height of 5.5 nm, in agreement with the expected thickness of a DPPC bilayer in the gel phase. The lowest layer has a thickness of approx. 1.5 - 2 nm and is too thin to be attributed to a bilayer. It is also thinner than a structured leaflet, which would be about half of a bilayer, suggesting that it corresponds to an unstructured DPPC monolayer directly on top of the $Si/SiO_x$ surface.

The pyramidal staking of the bilayers exposes extended flat regions of different thickness, allowing to study the change in their mechanical properties with increasing number of layers.

Figure 1B maps the value of the local Young's Modulus calculated from the Derjaguin-Mueller-Toporov model of indentation ($E_{DMT}$), providing values in the range between 0.1 and 1.8 MPa. The image shows a gradual decrease of $E_{DMT}$ while progressing from the bare Si/SiO$_x$ surface to the higher bilayer stacks. The relatively low force constant of the cantilever (approx. 1.4 N/m) and high Young modulus of Si (approx.130 GPa for the bare Si(111) surface) and SiO$_x$ (approx. 50 GPa) prevent the accurate determination of $E_{DMT}$ for the Si/SiOx surface due to the limited indentation. The measured values, in the range 1.0 - 1.8 GPa, are to be considered inaccurate and greatly underestimate the real value. The thickness and lower Young modulus of the single leaflet region in principle allow sufficient indentation for a more accurate estimate of the value of $E_{DMT}$. Nonetheless they provide values in the range 0.5 – 1.5 GPa, much higher than expected for a soft sample. Single and multiple bilayers provide $E_{DMT}$ values decreasing from 1 GPa to 200 MPa with increasing number of bilayers. The lower values are in the range expected for soft samples, showing that the presence of the substrate becomes progressively less noticeable with increasing sample thickness.

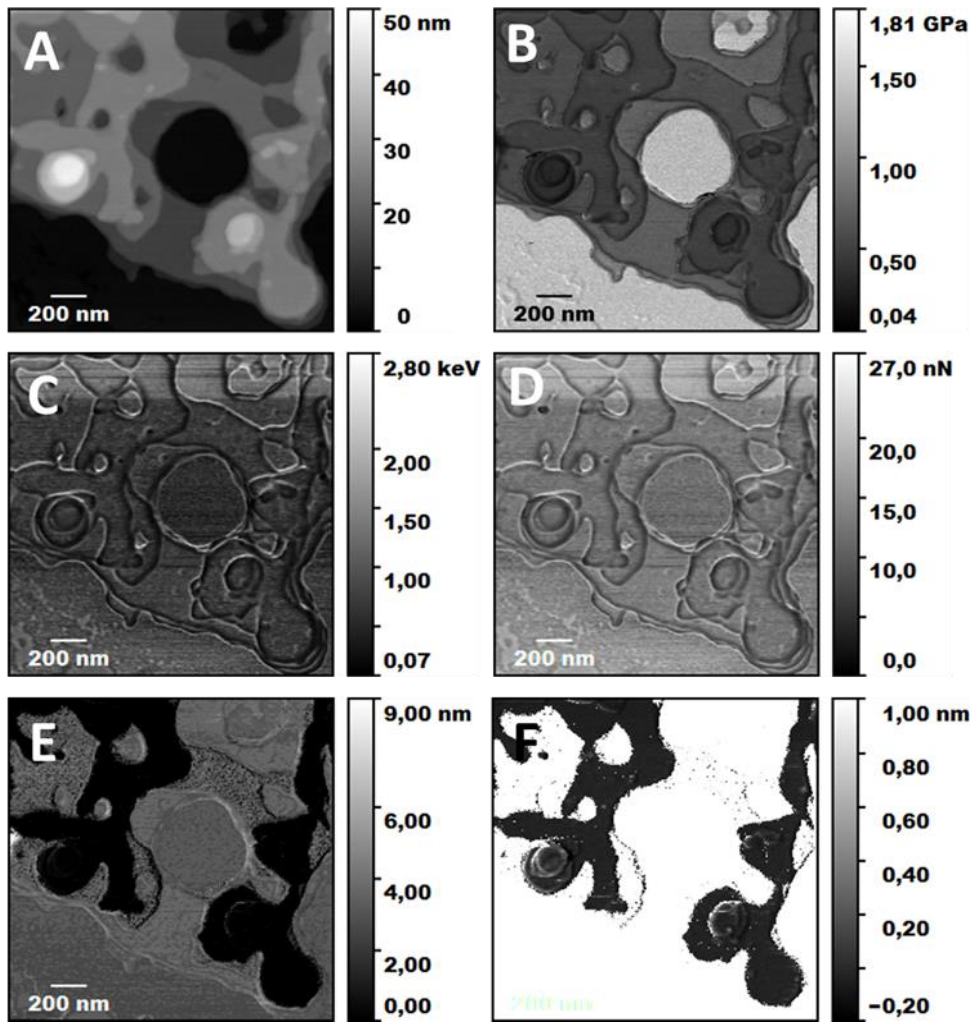

*Figure 2. QNM maps of dry DPPC multibilayer stacks (200 pN Setpoint). **A.** Z Sensor; **B.** Modulus; **C.** Dissipation; **D.** Adhesion; **E, F.** Deformation, with two different scale ranges.*

Figure 2C and Figure 2D show the dissipation and adhesion maps of the sample. Both dissipated energy and adhesion appear to be uniform on top of DPPC bilayers, and unaffected by the height of the sample and the number of underlying bilayers. However, both are noticeably higher in locations where the Si/SiO$_x$ surface is exposed to direct contact with the tip, presumably because of capillary interactions between tip and sample. A change in overall intensity of the signal is observed about fifty lines from the top of the map and can be attributed to a sudden change in tip-sample interactions, presumably because of the loss of or collection of a minor tip contaminant.

Figure 2E and Figure 2F map the deformation across the sample using two different scales. Surprisingly, deformation values appear in the 2 - 6 nm range for three or less bilayers. In contrast they fall in the 0 - 1 nm range for four bilayers or more.

No effect was observed on the structure and topography of the bilayers despite repeated scans with setpoint values ranging from 200 pN to 1800 pN (see supporting Figure 2). As noted previously, the only progressive change was reported in deformation values (see supporting Figure 2). The latter showed a variability that depended on the selected setpoint but also on the time of the measurements, suggesting an effect due to aging of the sample.

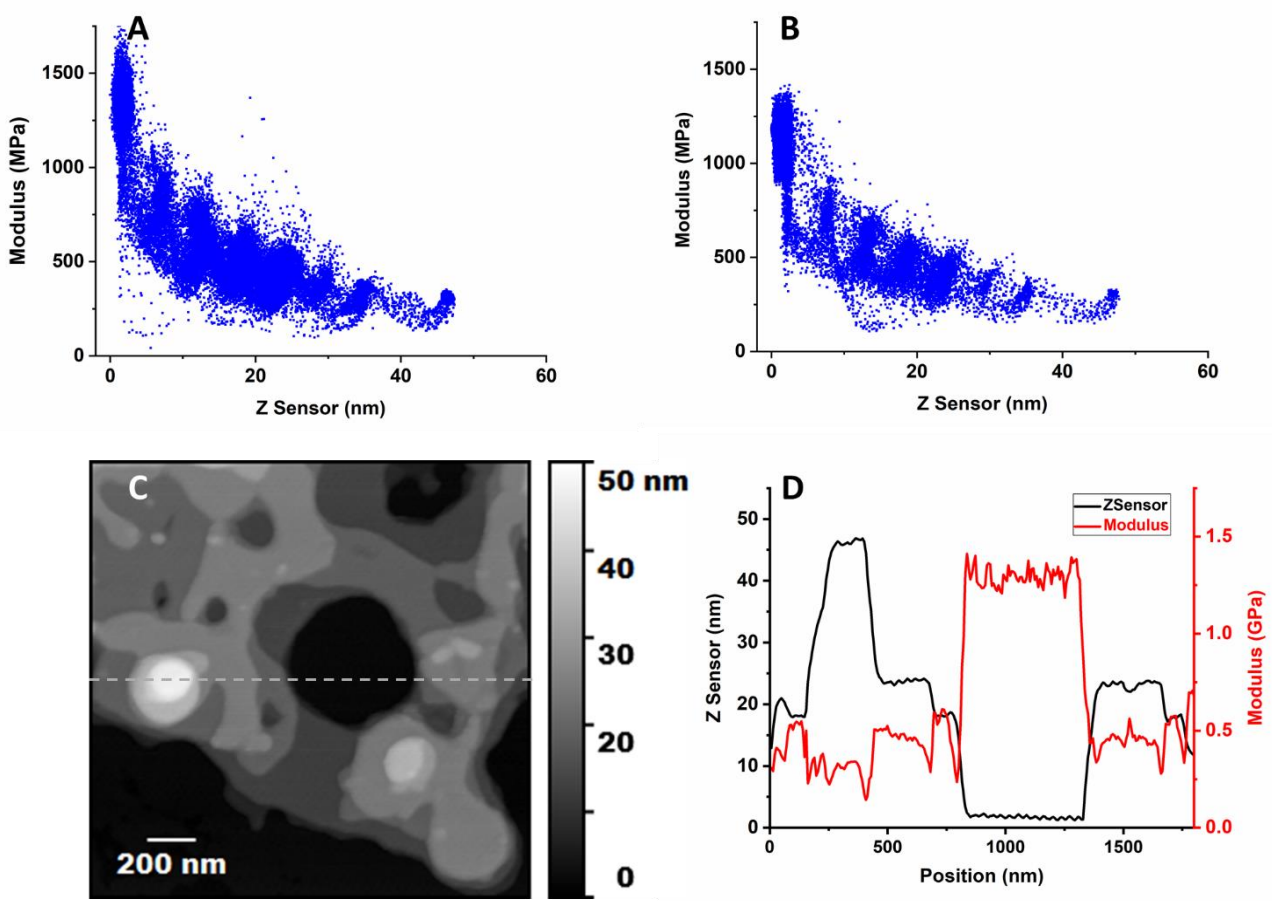

*Figure 3. Relationship between $E_{DMT}$ and Z value in a DPPC multibilayer stack. A. Correlation between the decrease of $E_{DMT}$ and the increase in Z value using a 200 pN setpoint. This negative linear*

*dependence stops above approx. 1 GPa, corresponding to regions of the Si/SiOx surface, where the measured value of the modulus is inaccurate. **B.** Correlation between the decrease of $E_{DMT}$ and the increase in Z value using a 1800 pN setpoint. **C.** Z Map of DPPC multibilayers. The dotted line shows the profile represented in **D**. Comparison of Z and E along the dotted line in panel B, showing the inverse relationship between the two.*

Figure 3 shows the inverse dependence between the height of the multibilayer, as represented by Z, and the value of $E_{DMT}$. Figures 3A and 3B plot the values of Z and $E_{DMT}$ for each point of a sample map (not shown), at 200 pN and 1800 pN setpoint values respectively. Below 1 GPa a clear trend is observed, corresponding to a decrease of $E_{DMT}$ for increasing Z. Above approximately 1 GPa the diagram flattens out, indicating to a high variability of $E_{DMT}$ for small changes in Z. The latter region corresponds to the naked Si/SiOx surface. Measurements of $E_{DMT}$ in this region are inaccurate with the tip in use, given the high modulus of Si(111) (>130 GPa). Careful observation of Figure 3A and Figure 3B shows that the data points form separate clusters, corresponding to stacks of 1 to 8 bilayers. A thin cluster corresponding to a monomolecular layer is also seen at the side of the diagram, between approx. 0.8 GPa and 1.0 GPa. The inverse relationship between $E_{DMT}$ and Z is confirmed in the profile comparison of Figure 3C. The latter also highlights that a large drop in the value of $E_{DMT}$ is observed at the edges of the stacks, which is responsible for the large spread of $E_{DMT}$ values in the clusters of Figure 3A and Figure 3B. The measurements were repeated at different setpoints to probe the effect of varying the peak force. Comparison of the maxima in the distribution of Z values (not shown), corresponding to the plateaus of the stack, indicates small or negligible changes in height. Maxima are observed at 2.2 nm (DPPC monolayer), 7.8 nm ($1^{st}$ bilayer), 13.1 ($2^{nd}$ bilayer), 18.3 ($3^{rd}$ bilayer), 24.3 ($4^{th}$ bilayer), 30.0 ($5^{th}$ bilayer), 35.8 ($6^{th}$ bilayer), 46.8 nm ($8^{th}$ bilayer) with 200 pN and at 2.0 nm (DPPC monolayer), 7.7 nm ($1^{st}$ bilayer), 13.0 nm ($2^{nd}$ bilayer), 18.3 nm ($3^{rd}$ bilayer), 23.9 nm ($4^{th}$ bilayer), 29.1 nm ($5^{th}$ bilayer), 35.1 nm ($6^{th}$ bilayer), 46.8 nm ($8^{th}$ bilayer) with 180 pN. The results in Figure 3A and Figure 3B, at 200 pN and 1800 pN, also show that a small decrease of the average value of $E_{DMT}$ can be appreciated when using higher peak force. In addition, measurements at higher peak force display a reduced spread, presumably because of the improved precision of the measurement.

To obtain more accurate values of $E_{DMT}$ for the flat region of each stack, without interference from the edge regions, several points along selected lines in Figure 3C were averaged. The result is shown in Table I. Table I also compares the values of $E_{DMT}$ obtained from two different models of tip-sample interaction, Hertz, Sneddon and DMT (indicated as $E_{Hertz}$, $E_{Sneddon}$, $E_{DMT}$). Hertz and DMT models both involve indentation by a spherical tip and give comparable values of E. In contrast, Sneddon model assumes indentation by a sharp conical tip, leading to much larger values of E. Despite differences in absolute numbers, the inverse relationship between E and Z is confirmed for all models. The observed decreasing trend for E is interrupted only at the transition between the $3^{rd}$ and $4^{th}$ bilayer, corresponding to the discontinuity observed in deformation maps.

***Table I. Average Values of Young Modulus E of DPPC Multilayers for Different Models of Indentation**. The standard deviation of the measurement is given in brackets. 200 pN peak force setpoint.*

|               | $E_{Hertz}$ (MPa) | $E_{Sneddon}$ (MPa) | $E_{DMT}$ (MPa) | Z Sensor (nm) |
|---------------|-------------------|---------------------|-----------------|---------------|
| **Substrate**     | 1700 (144)        | 35600 (3520)        | 1290 (35)       | 1.76 (0.20)   |
| **Leaflet**       | 1550 (102)        | 26800 (2830)        | 1370 (98)       | 1.78 (0.37)   |
| **1st Bilayer**   |                   |                     | 1000 (46)       | 5.90 (0.21)   |
| **2nd Bilayer**   | 851 (67)          | 12600 (1410)        | 686 (32)        | 12.2 (0.18)   |
| **3rd Bilayer**   | 521 (20)          | 7240 (442)          | 474 (26)        | 18.5 (0.14)   |
| **4th Bilayer**   | 599 (31)          | 7700 (442)          | 475 (13)        | 23.8 (0.13)   |
| **5th Bilayer**   | 485 (23)          | 7040 (568)          | 360 (32)        | 27.4 (0.28)   |
| **6th Bilayer**   |                   |                     | 346 (50)        | 34.1 (0.81)   |
| **8th Top Bilayer** | 361 (14)        | 4281 (212)          | 325 (14)        | 46.2 (0.35)   |

Liposomes were deposited on Si/SiO$_x$ substrates by direct opening from an aqueous solution. The samples were monitored using Peak Force QNM during the first few hours, revealing the formation of patches of single bilayers and of scattered liposomes. Only few double bilayers were observed, which appeared formed by the recent opening of a liposome on top of a single bilayer. One such example is shown in Figure 4, reporting topography and mechanical properties. As for the case of dry bilayer stacks, $E_{DMT}$ decreases in the core upper bilayer compared to the lower bilayer. The upper layers also show higher deformation under conditions of constant peak force. In contrast to dry bilayers, higher $E_{DMT}$ values are observed along the edges of the patches in Figure 4, together with lower deformation, indicating that edges are more rigid than the core. The latter observation is not always reproduced, and some patches (not shown) display the same $E_{DMT}$ value to the very edge.

Measured values of the adhesion force are about two orders of magnitude higher in an aqueous environment, as already reported.[4] In contrast to dry bilayers, where adhesion forces show little or no dependence from the Z Scanner position, a lower adhesion is clearly displayed on the upper bilayer of aqueous stacks.

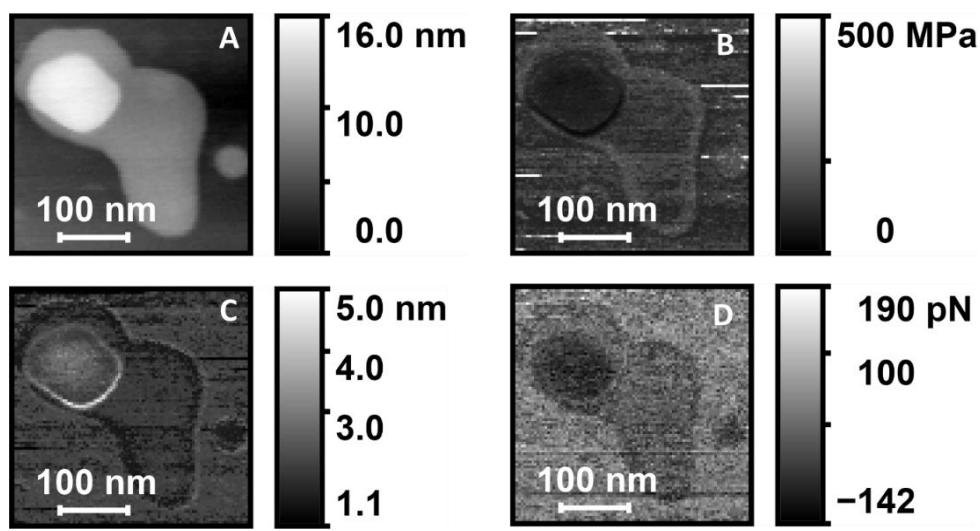

*Figure 4. QNM maps of aqueous DPPC multibilayer. **A**. Z Sensor; **B**. DMT Modulus; **C** Deformation; **D** Adhesion.*

Figure 5 maps the mechanical properties of DPPC liposomes freshly deposited on the Si/SiOx surface. The liposome is about 100 nm in diameter and 50 nm thick, with a round flattened top, as shown in Supporting Figure 2. A barely perceptible rim surrounds the liposome, presumably formed by adhesion to the surface, where the surface curvature makes an inflection. Figure 5B shows the measured $E_{DMT}$ value across the sample. The scale has been reduced to allow discriminating regions with different mechanical properties within the liposome. Values of $E_{DMT}$ are about one order of magnitude lower than for the supported DPPC bilayers observed in the same sample. The highest values of $E_{DMT}$ correspond to flatter regions, namely the rim and the top of the flattened particle (approx. 10 MPa). The lowest values are observed at the region of maximum curvature of the particle. Deformation values follow the complementary trend to $E_{DMT}$, with lower values observed for flatter regions and higher values for curved regions. In contrast, adhesion values appear to decrease progressively at higher Z values.

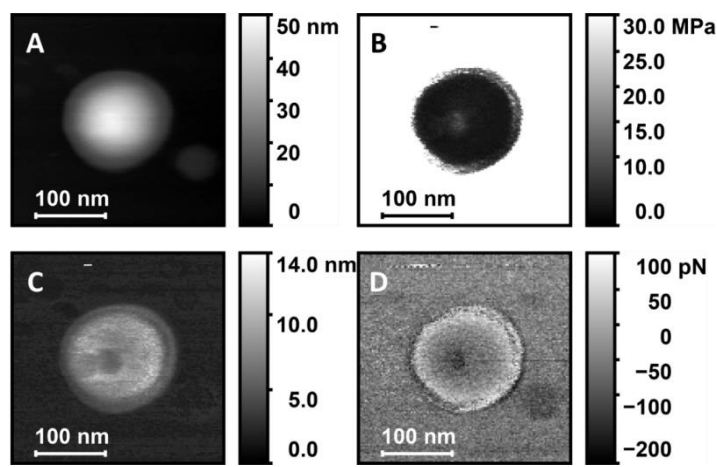

*Figure 5. Mechanical Properties of Aqueous DPPC Liposomes on the Si/SiO$_x$ surface. **A.** Z Sensor; **B.** DMT Modulus; **C.** Deformation; **D.** Adhesion.*

**Discussion**

We report that the presence of the rigid substrate does affect the Young modulus of bilayers in a stack and this effect is perceived as far as six stacked bilayers (approx. 35 nm). Beyond this limit the modulus appears to converge to a stable value. We take the limiting value measured for E of DPPC bilayers, approx. 300 -350 MPa, as the value for E when substrate effects have been minimized. This is comparable, to the values already reported by some authors for supported gel-phase DPPC bilayers. [5]

Comparison of the Young modulus extracted from different indentation models confirms the decreasing trend with increasing number of bilayers, despite the different values reported for individual models. Furthermore, all models show a discontinuity in this trend between the same bilayers that border the two regions of different deformation. Transitioning between the 3$^{rd}$ and the 4$^{th}$ bilayer, the Young modulus does not decrease, whereas deformation drops from about 5 nm to less than 1 nm (Figure 1 and Table I). Overall, the behavior indicates the presence of a sharp change in the mechanical properties of the system at this threshold.

The values of $E_{DMT}$ and $E_{Hertz}$ are always $E_{DMT} < E_{Hertz}$, with the difference being given by the adhesion term included in the DMT model. Adhesion appears to be approximately constant on dry flat DPPC surfaces and does not affect the dependence of E from bilayer thickness. The values of $E_{Sneddon}$, while following the same trend as the values form other models, is unusually higher. The reason for the discrepancy is presently unclear.

The values of $E_{DMT}$ and $E_{Hertz}$ measured in these experiments are at least three orders of magnitude larger than the one reported from force measurements of intact cells, in the range of 1-100kPa [6], and about ten times larger than the equivalent values reported for most other supported bilayers in the liquid disordered phase. [7] The discrepancy can be ascribed to the specific properties of DPPC, which is in the liquid ordered phase below 40°C. The relatively high values of E measured even when probing thicker bilayer stacks, are consistent with a highly ordered structure of gel phase DPPC. This is consistent with the structure of DPPC. Studies of supported DPPC bilayers by nanoFTIR have already indicated that the bilayer has a highly ordered structure, with alkyl chains closely packed in the all-trans configuration and the headgroups regularly oriented along the surface. [8] X-ray diffraction also suggests that in the DPPC the structure display crystalline lattices for the headgroups and hydrocarbon chains. In this respect it may be inappropriate to describe the gel phase of DPPC as a liquid ordered phase, while the two terms are often used interchangeably for other phospholipid. [9] This is in sharp contrast to the structure of biological membranes, which in general are in the lipid disordered state and display a fluid structure, allowing high molecular mobility. [10] Despite the difference, the DPPC system described in the present work can still be a relevant nano mechanical (although not compositional) model for the highly ordered phospholipid domains proposed to coexist within the fluid portions of cellular membranes. [11] These results bear implication for the interpretation of the mechanical properties of single cells, which have been mostly ascribed to changes in the organization of the cytoskeleton, whereas the role of lipids has received comparatively little attention. The high values reported for the Young modulus of ordered lipid phases suggests that this approach may need a revision.

Aqueous liposomes provide a model for a DPPC bilayer in an aqueous environment, similar to the physiological conditions of biological membranes. In particular, the upper portion of the bilayer is separated from the substrate by the aqueous core of the liposome and provides a closer approximation to a cellular membrane than a flat supported bilayer. Young's modulus for the liposome surface ranges from 2 MPa to 10 MPa, one to two orders of magnitude lower than the lowest value observed for the top of flat bilayer stacks. The lower values are observed on the sides of the liposome, where the curvature is maximal. In contrast, the highest value is observed at the top of the liposome, where the membrane is horizontal.

Even for a liposome membrane, the values of E are at least one order of magnitude higher than the ones reported for intact cells. The residual difference may be partially attributable to the specific properties of gel phase DPPC bilayers. However, it must also account for the large compositional difference of cellular membranes, which include complex lipid and biopolymer mixtures.

The observation that the modulus of phospholipid bilayers decreases in convex regions and increases in concave regions has already been reported for ripple phase bilayers. [12] The deformation of the liposome is also a function of curvature, with the highest deformation values, in excess of 10 nm, observed on the convex side surface (Figure 5). The specific mechanical properties of curved surfaces may be ascribed to differences in structures between flat and curved phospholipid surfaces, which have been related to differences in the phase of the systems. [13] The contact rim and the top of the liposome display 5 nm or less deformation. Overall, the mechanical properties of the liposome suggest an analogy with an architectural archway. The force applied on top of the keystone element is transmitted symmetrically via the side voussoirs to the supports, thus ensuring the stability of the whole structure under heavy load.

Differences in the mechanical properties of bilayer edges have already been observed in tapping mode AFM. In agreement, we now show that, in the case of dry DPPC bilayers, edges generally display lower modulus. However, such decrease in stiffness appears to have a different origin than the curvature of liposome membranes. Topography profiles of the edges indicate that the bilayer appears continuous until it drops with a sharp edge, without any obvious curvature within the resolution allowed by tip size. One possible interpretation is that they arise from the reorganization of phospholipid molecules at the edge following compression, corresponding to lower E and reduced adhesion and dissipation. The edge-core difference could also be ascribed to a difference in Poisson parameter due to the possibility for the molecules in this region to be displaced outside of the bilayer.

Different spectroscopic properties of DPPC bilayer edges have been reported using nanoFTIR.[8] However, it is presently unclear whether the nanoFTIR response is determined by a cross-talk between mechanical and spectroscopic properties or by structural differences at the molecular level that affect both spectroscopic and mechanical properties. Further comparative studies are necessary to clarify this point.

Adhesion forces at the surface of aqueous bilayers are two orders of magnitude lower than adhesion forces in dry bilayers, in agreement with past experimental observations. The magnitude of this difference has been explained by the attractive forces due the formation of a water meniscus in dry sample in humid environment. In agreement with this interpretation, during the experiments on dry bilayers environmental humidity was higher than 20-30%, the threshold at which adhesive forces show a sharp increase attributed to meniscus formation. [4] However, the meniscus model fails to account for the decreased adhesion of aqueous upper bilayers reported in this work (Figure 1), suggesting that a revision may be needed. Adhesion on the surface of a supported liposome also shows progressively decreasing values with increasing Z value and is minimal at the top of the liposome. The Z dependence of adhesion indicates that a non-contact long range interaction is involved, possibly an electrostatic interaction mediated by the electrolyte in solution.[14]

A remarkable result of this work is the observation of two different regions in deformation maps. A region with high deformation, up to 6 nm, is observed for the lower levels of the stack, whereas a region of lower deformation, 1-2 nm, is observed for the higher bilayers. The boundary level between the two is determined by the applied peak force setpoint. It moves to progressively lower height with increasing peak force and disappears at 1800 pN. In parallel, a decrease in the value of E is observed, together with a reduced spread of E values. Because of the apparent dependence from sample aging, we tentatively propose that this effect could be attributed to the presence of a hydration layer at the headgroups of DPPC. Hydration water increases the plasticity of the bilayer stack by its extended hydrogen bonding network, reducing the interaction between headgroups, and allowing the bilayers to more easily flex under pressure. Loss of hydration water from sample aging or the application of a higher setpoint results in decreased plasticity and increased elasticity of the sample, leading to smaller deformation.

**Conclusions**

We have described the mechanical properties of nanoscale structures based on DPPC bilayers, including stacked bilayers and liposomes, in an aqueous environment and in air. We have used the properties of stacks of bilayers to probe the distance dependence of the substrate effect on the Young modulus of the bilayers and report a convergence to a value of circa 300-350 MPa after about 35 nm (6 bilayers). The high value of the modulus supports the description of DPPC bilayers as rigid ordered structures. We compare this with the value of 10 MPa observed for the aqueous upper bilayer of liposomes. We observe a dependence of the mechanical properties of the bilayers on surface curvature, with the curved surfaces of liposomes displaying a lower Young modulus and higher deformability. Different mechanical properties are observed for the edges of dry bilayers relative to the core of the bilayer, while the difference is smaller and inconsistent for aqueous bilayers. The behavior of adhesion forces appears to be unusual in aqueous samples, displaying a lower value at higher distance between the DPPC surface and the substrate. Finally, we report the existence of a threshold level in bilayer stacks between a region of higher plasticity, at the base of the stack, and a region of lower plasticity, at the top of the stack, which we ascribe to different levels of interstitial hydration between the headgroups of adjoining bilayers. The presence of a thin aqueous layer intercalated between the molecular-scale structure of the phospholipid headgroups provides for a unique two-dimensional architecture, allowing the simultaneous interplay of atomic and bulk properties, which can give rise to interesting mechanical, thermal, optical and electrical properties.

**Acknowledgments**

This research was supported by an OPUS16 grant to LQ from the National Science Center Poland under contract 2018/31/B/NZ1/01345. L.Q. and K.B. were also supported by the European Union's Horizon 2020 research and innovation programme under the Marie Skłodowska-Curie grant agreement no. 665778, managed by the National Science Center Poland under POLONEZ grant to LQ contract 2016/21/P/ST4/01321.

LQ is grateful to Dr. Michał Szuwarzyński, AGH Kraków, and Dr. Karol Wolski, for advise adn discussion on AFM usage.

**Supporting Information: Nanomechanical Characterization of Bilayers of 1,2-dipalmitoyl-sn-glycero-3-phosphocholine.**

**1.0 Setpoint dependence of deformation**

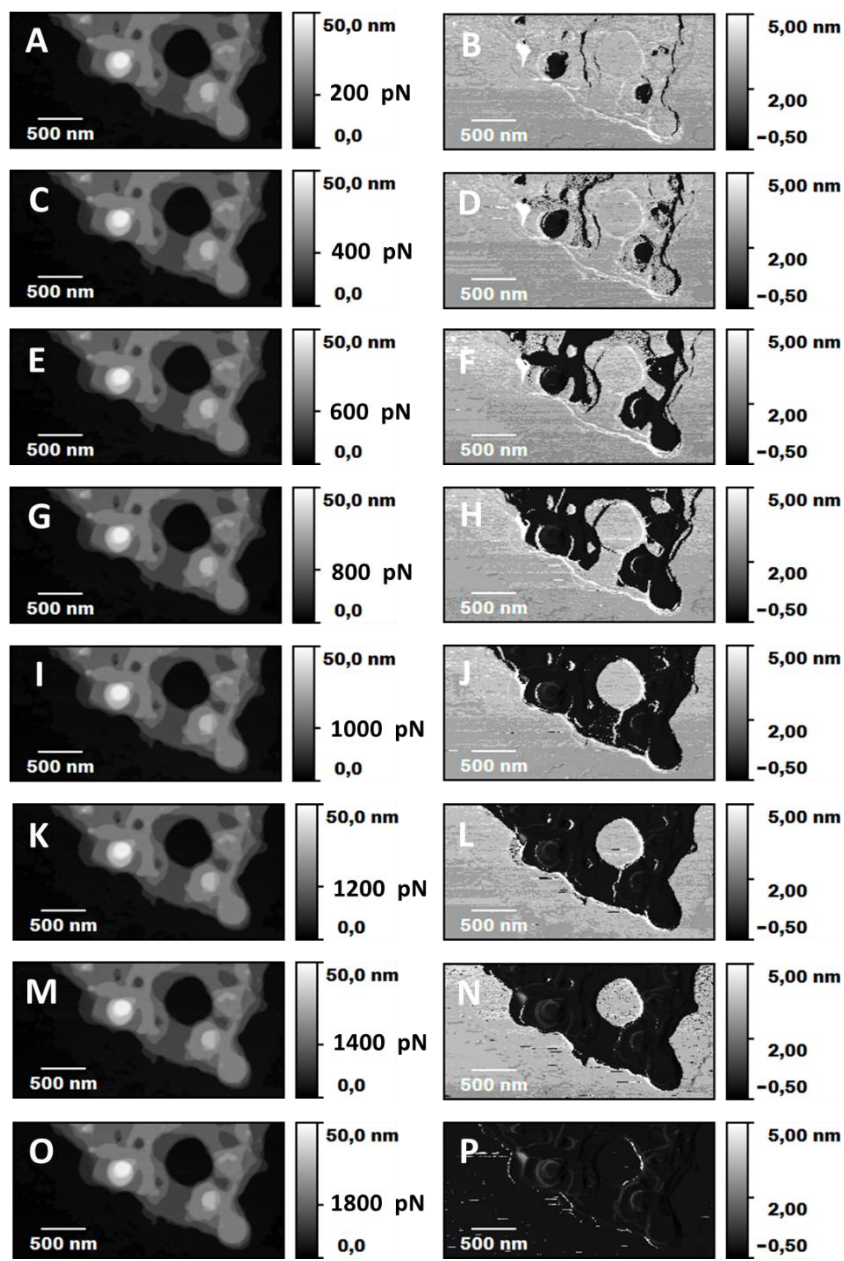

*Supporting Figure 1. Z sensor and Deformation maps of bilayer stacks as a function of setpoint.*

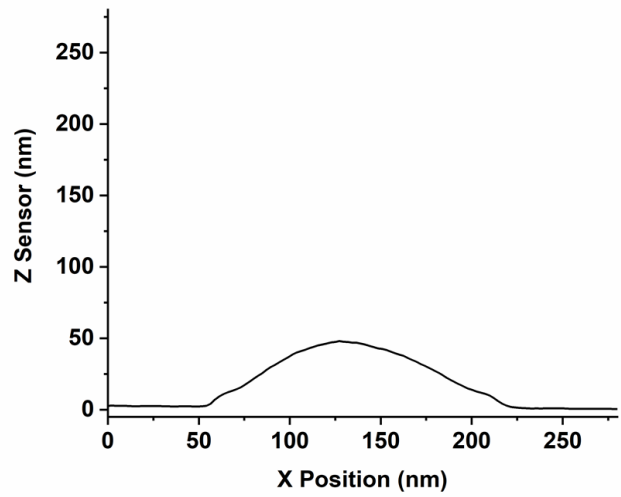

***Supporting Figure 2.*** *Profile of the aqueous supported liposome mapped in Figure 5 of the main text.*